\documentclass[
prl,
reprint,
twocolumn,
 linenumbers,	
amssymb, amsmath, aps, showpacs, nofootinbib, superscriptaddress]{revtex4-1}

\usepackage{subfigure}
\usepackage{graphicx}

\usepackage{natbib}
\usepackage{captcont}
\usepackage[final,		
	colorlinks=true,			
	linktocpage,		
	linkcolor=blue,		
	citecolor=red,		
	urlcolor=blue,		
	breaklinks=true,	
	]{hyperref}
\usepackage[all]{hypcap}	

\usepackage{breakurl}

\begin{document}
\title{Measurement of separate cosmic-ray electron and positron spectra with the Fermi Large Area Telescope}

\author{M.~Ackermann}
\email{markus.ackermann@desy.de}
\affiliation{Deutsches Elektronen Synchrotron DESY, D-15738 Zeuthen, Germany}
\author{M.~Ajello}
\affiliation{W. W. Hansen Experimental Physics Laboratory, Kavli Institute for Particle Astrophysics and Cosmology, Department of Physics and SLAC National Accelerator Laboratory, Stanford University, Stanford, CA 94305, USA}
\author{A.~Allafort}
\affiliation{W. W. Hansen Experimental Physics Laboratory, Kavli Institute for Particle Astrophysics and Cosmology, Department of Physics and SLAC National Accelerator Laboratory, Stanford University, Stanford, CA 94305, USA}

\author{W.~B.~Atwood}
\affiliation{Santa Cruz Institute for Particle Physics, Department of Physics and Department of Astronomy and Astrophysics, University of California at Santa Cruz, Santa Cruz, CA 95064, USA}

\author{L.~Baldini}
\affiliation{Istituto Nazionale di Fisica Nucleare, Sezione di Pisa, I-56127 Pisa, Italy}
\author{G.~Barbiellini}
\affiliation{Istituto Nazionale di Fisica Nucleare, Sezione di Trieste, I-34127 Trieste, Italy}
\affiliation{Dipartimento di Fisica, Universit\`a di Trieste, I-34127 Trieste, Italy}
\author{D.~Bastieri}
\affiliation{Istituto Nazionale di Fisica Nucleare, Sezione di Padova, I-35131 Padova, Italy}
\affiliation{Dipartimento di Fisica ``G. Galilei", Universit\`a di Padova, I-35131 Padova, Italy}
\author{K.~Bechtol}
\affiliation{W. W. Hansen Experimental Physics Laboratory, Kavli Institute for Particle Astrophysics and Cosmology, Department of Physics and SLAC National Accelerator Laboratory, Stanford University, Stanford, CA 94305, USA}
\author{R.~Bellazzini}
\affiliation{Istituto Nazionale di Fisica Nucleare, Sezione di Pisa, I-56127 Pisa, Italy}
\author{B.~Berenji}
\affiliation{W. W. Hansen Experimental Physics Laboratory, Kavli Institute for Particle Astrophysics and Cosmology, Department of Physics and SLAC National Accelerator Laboratory, Stanford University, Stanford, CA 94305, USA}
% ...


\begin{abstract}
We measured separate cosmic-ray electron and positron spectra with the Fermi Large Area Telescope.  Because the instrument does not have an onboard magnet, we distinguish the two species by exploiting the Earth's shadow, which is offset in opposite directions for opposite charges due to the Earth's magnetic field.  We estimate and subtract the cosmic-ray proton background using two different methods that produce consistent results.  We report the electron-only spectrum, the positron-only spectrum, and the positron fraction between 20~GeV and 200~GeV.  We confirm that the fraction rises with energy in the 20--100~GeV range.  The three new spectral points between 100 and 200 GeV are consistent with a fraction that is continuing to rise with energy.
\end{abstract}

\pacs{96.50.S-, 95.35.+d, 95.85.Ry, 98.70.Sa}

\date{\today}%
\maketitle


\emph{Introduction.}  Positrons were discovered in cosmic rays (CRs) in 1964~\cite{DeShong1964}.  Subsequent measurements through the 1960s and 1970s~\cite{Fanselow1969, Daugherty1975} showed that the positron fraction, $e^+/(e^+ + e^-)$, decreases with energy up to $\sim$10~GeV.  See~\cite{Protheroe1982} and references therein for a summary of these early experiments.  Balloon payloads through the 1980s measured CR electrons and positrons with inconsistent results~\cite[][and references therein]{Muller1990}.  Starting in the mid-1990s, HEAT~\cite{HEAT1995,HEAT1997,HEAT1998,HEAT2001,HEAT2004}, CAPRICE~\cite{CAPRICE1996,CAPRICE2000}, and TS93~\cite{TS93} measured the CR positron spectrum and fraction up to $\sim 50$~GeV, indicating a predominantly decreasing positron fraction with increasing energy.  However, a small excess in the positron fraction above $\approx 7$~GeV was detected by HEAT and also seen in CAPRICE data, as well as by AMS-01~\cite{Aguilar2007}.  Recently, the PAMELA instrument has measured a positron fraction that increases with energy above $\sim$10~GeV~\cite{Adriani2009,Pamela2} with high precision, confirming the indications seen in the earlier data.

The best established mechanism for producing CR positrons is secondary production: CR nuclei interact inelastically with interstellar gas, producing charged pions that decay to positrons, electrons, and neutrinos.  However, this process results in a positron fraction that decreases with energy~\cite{Protheroe1982,Moskalenko1998}.  The origin of the rising positron fraction at high energy is unknown and has been ascribed to a variety of mechanisms including pulsars, CRs interacting with giant molecular clouds, and dark matter.  See~\cite{Fan2010,Porter2011} for recent reviews.

The Large Area Telescope (LAT) is a pair-conversion gamma-ray telescope onboard the Fermi Gamma-ray Space Telescope satellite.  It has been used to measure the combined CR electron and positron spectrum from 7~GeV to 1~TeV~\cite{LATCRE, LATCRE1}.  The LAT does not have a magnet for charge separation.  However, as pioneered by~\cite{Daniel1965} and~\cite{Muller1987}, the geomagnetic field can also be used to separate the two species without an onboard magnet.  M\"{u}ller and Tang~\cite{Muller1987} used the difference in geomagnetic cutoff for positrons and electrons from the east and west to determine the positron fraction between 10~GeV and 20~GeV.  As reported below, we used the shadow imposed by the Earth and its offset direction for electrons and positrons due to the geomagnetic field, to separately measure the spectra of CR electrons and positrons from 20~GeV to 200~GeV.  In this energy range, the 68\% containment radius of the LAT point-spread function is 0.1$^\circ$ or better and the energy resolution is 8\% or better.


\emph{Region selection and exposure calculation.}  The Earth's magnetic field significantly affects the CR distribution in near-Earth space.  At energies below $\sim$10~GeV, a significant fraction of the incoming particles are deflected back to interplanetary space by the magnetic field (``geomagnetic cutoff'').  The exact value of the geomagnetic cutoff rigidity depends on the detector position and viewing angle.  In addition to the geomagnetic cutoff effect, the Earth blocks trajectories for particles of certain rigidities and directions while allowing other trajectories.  This results in a different rate of CRs from the east than the west (the ``east-west effect'')~\cite{Rossi1930, Johnson1933, Alvarez1933}.

\begin{figure}[]
\includegraphics[width=\linewidth]{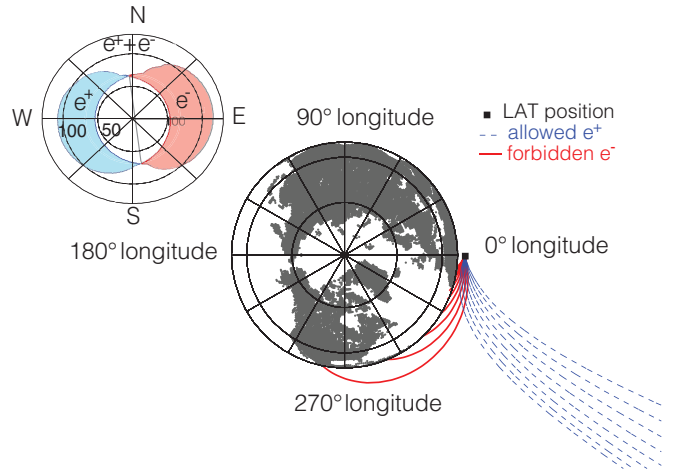}
\caption{Examples of calculated electron (red) and positron (blue) trajectories arriving at the detector, for 28~GeV particles arriving within the Equatorial plane (viewed from the North pole).  Forbidden trajectories are solid and allowed trajectories are dashed.  Inset: the three selection regions (electron-only, positron-only, and both-allowed) for the same particle energy and spacecraft position as the trajectory traces (viewed from the instrument position in the Equatorial plane).}
\label{rayTrace}
\end{figure}

Figure~\ref{rayTrace} shows example trajectories for electrons and positrons.  Positive charges propagating toward the east are curved outward, while negative charges are curved inward toward the Earth (Figure~\ref{rayTrace}).  This results in a region of particle directions from which positrons can arrive, while electrons are blocked by the Earth.  At each particle rigidity there is a region to the west from which positrons are allowed and electrons are forbidden.  There is a corresponding region to the east from which electrons are allowed and positrons are forbidden.  The precise size and shape of these regions depend on the particle rigidity and instrument location.

We used a high-precision geomagnetic field model (the 2010 epoch of the 11th version of the International Geomagnetic Reference Field~\cite{igrf}) and a publicly available code~\cite{tracer} to trace charged particle trajectories in the magnetic field and determine allowed vs. forbidden regions for each species.  We previously used the same magnetic field model and tracer code to perform a precise comparison between predicted and measured geomagnetic cutoff rigidities for the Fermi LAT orbit, finding that the tracer code accurately predicts the geographical distribution of the geomagnetic cutoff~\cite{Melissa}.  We used a static 2010 model for all of the data analyzed here, which were recorded between June 2008 through April 2011.

Each particle trajectory is traced backward from the spacecraft until it reaches 20 Earth radii from the Earth center or reaches the Earth's atmosphere, which we approximate with a 60~km thickness (Figure~\ref{rayTrace}).  If the trajectory reaches 20 Earth radii, it is an allowed trajectory.  If it reaches the atmosphere, it is a forbidden trajectory.  We calculate electron-only, positron-only, and both-allowed (control) regions for each 30~s time step using the instantaneous spacecraft latitude and longitude and the nominal orbital altitude of 565~km.  The regions are determined for each energy bin, with 10 logarithmically spaced energy bins spanning 20--200~GeV.  The 30~s time step (in which the spacecraft travels $\sim2^\circ$ longitude) is sufficient to achieve a finely sampled distribution of instantaneous regions and exposures.  Although we use binned position data for the exposure calculation, we use the instantaneous spacecraft position at the time of each event to determine which region it lies in, so the event selection is not affected by the finite step size.  

We define the ``deflected horizon'' (inset of Figure~\ref{rayTrace}) to be the curve that separates the allowed and forbidden regions for a particular energy, charge, and spacecraft position.  The curve represents the position of the Earth horizon as ``deflected'' by the geomagnetic field, with deflection occurring in one direction for electrons and in the opposite direction for positrons.  At 20~GeV the curves differ significantly from the actual Earth horizon.  At higher energies, the curves asymptotically approach the undeflected horizon.  The positron-only region is defined to be the region above the positron horizon and below the electron horizon.  A corresponding definition is used for the electron-only region.  The region above both instantaneous horizon curves and below a nadir angle of 130$^\circ$ is the control region, where both species are allowed.  The control region is truncated at 130$^\circ$ because the additional statistics gained using a larger region are unnecessary for this analysis and require more processing time.  The regions vary with spacecraft position and particle energy.  

We determine the integrated exposure (cm$^2$ sr s) for each of the three regions by integrating the energy- and direction-dependent effective area of the instrument over the solid angle of each region and over the livetime of the data set.  The numerical integration is performed with 0.5$^\circ$ pixels.  For each particle, we use the reconstructed energy and actual spacecraft position at the time of the event to determine the deflected electron and positron horizon curves.  We then calculate the nadir angle difference between the particle arrival direction and the deflected horizon: $D_{\pm} = N - N_{\pm}$, where $N$ is the nadir angle from which the event arrived and $N_{\pm}$ is the nadir angle of the positron (electron) horizon, evaluated at the azimuth direction from which the event arrived.  As described below, we use $D_{\pm}$ to select one set of CR electrons and another set of CR positrons and to remove atmospheric electrons and positrons.

When a CR nucleus interacts in the Earth's atmosphere, it produces a shower of secondary particles including gamma rays, electrons, and positrons.  At the high energies considered here, the secondaries are closely collimated along the direction of the primary.  The limb of the atmosphere is therefore visible in gamma rays as a bright, thin (1-2$^\circ$ wide) ring of emission.~\cite{FermiAlbedo}.  The peak emission is produced at a nadir angle of 68$^\circ$, which is 1.3$^\circ$ ($\sim$60~km) above the Earth surface.

\begin{figure}[]
\includegraphics[width=\linewidth]{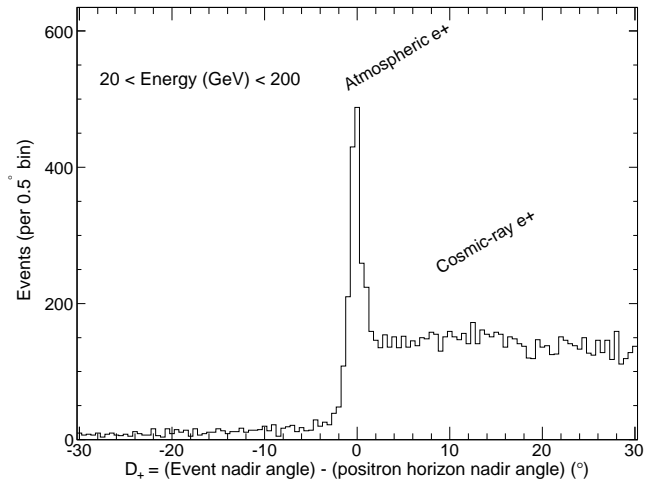}
\caption{Distribution of $D_+$, for events with $D_-<-2^\circ$.  The spike at $D_+=0^\circ$ is due to atmospheric positrons.  The events with $D_+>0^\circ$ are CR positrons plus residual proton background events.  The events with $D_+<0^\circ$ are residual background events.}
\label{limbSpike}
\end{figure}

Atmospheric electrons and positrons are produced with angular distributions similar to that of atmospheric gamma rays.  Unlike the gamma rays, they are deflected by the geomagnetic field before reaching the LAT and arrive from the direction of the deflected horizon curves rather than the actual Earth horizon.  After accounting for the deflection, however, they are also detected as a 1-2$^\circ$~wide ring of emission (Figure~\ref{limbSpike}).  Like the gamma-ray emission, the peak atmospheric positron emission seen by the LAT is produced at an impact parameter of $\sim$60~km above the Earth surface.  The detection of this population, and its narrow distribution at $D_+=0$, confirms that the magnetic field model and particle tracer code are accurate.  As a cross check, we also ran the trajectory tracer code with the 1995 magnetic field model.  Changing from the 2010 to the 1995 model, $D_{\pm}$ changes by less than 1$^\circ$ for most events, and the distribution shown in Figure~\ref{limbSpike} changes negligibly.

To remove atmospheric electrons and positrons, which are concentrated near $D_+=0$ and $D_-=0$, we trim the edges of our selection regions by a ``trim width'' $W$ in nadir angle.  We evaluated the residual contamination by atmospheric electrons and positrons by performing the full positron flux measurement as a function of $W$.  The contamination decreases significantly from $W=0^\circ$ to $W=2^\circ$ and is smaller than statistical fluctuations for $W\ge4^\circ$.  The contamination is $<$10\% for $W=2^\circ$ and $<$3\% for $W=4^\circ$.  Below 100~GeV we trim by 4$^\circ$.  Above 100~GeV we trim by 2$^\circ$; the systematic effect from residual secondaries with this less conservative trimming is larger (10\%) but still smaller than the statistical uncertainty in this energy range.  As explained below, we have included these estimates of the residual atmospheric contamination in the systematic uncertainties.  This effect is included for the positron flux but is negligible for the larger electron flux: negligible variation with $W$ is seen in the electron flux.


\emph{Run and event selection.}  Data collected in the LAT standard mode of operation (survey mode) are not suitable for this analysis.  In survey mode, the LAT points away from the Earth, keeping the electron-only and positron-only regions (which are centered on the Earth horizon) outside of or at the edge of the field of view.  However, the LAT has observed in special modes with the horizon in the field of view.  Direct limb-pointed observations have been performed for instrument commissioning and calibration~\cite{FermiAlbedo}.  Pointed observations have also been performed in which a particular celestial direction is tracked for several hours or several days.  During these observations the target is tracked until it is occulted by the Earth, during which time the spacecraft tracks along the limb of the Earth, with a 50$^\circ$ offset toward zenith, until the target re-emerges from occultation.  We used 775 runs from these observation categories with a total observation time of 39.0 days.  A LAT run is an interval of data taking that typically spans one orbit.

The recorded events are dominated by a background of CR protons and heavier nuclei.  Here we used the same particle classification technique that we developed to measure the combined $e^+ + e^-$ spectrum~\cite{LATCRE, LATCRE1} to separate CR leptons from this background.  We use events in the 20--200~GeV range.  Below 20~GeV the electron/positron sensitivity is small due to the event filter applied onboard the LAT: any event that deposits 20~GeV in the calorimeter passes this filter, but below this energy charged particles are significantly suppressed in favor of gamma rays~\cite{LATCRE1}.  Above 200~GeV, the electron-only and positron-only regions become very small; this combined with the steeply falling spectra makes the number of detected events prohibitively small.


\emph{Background subtraction.}  The CR hadron flux (predominantly protons) between 20~GeV and 200~GeV is 2--3 (depending on energy) orders of magnitude larger than the electron flux and 3--4 orders of magnitude larger than the positron flux~\cite{Strong2009}.  Our lepton/hadron separation criteria suppress the hadron background to a level that is between 4\% and 20\% (depending on energy) of the combined electron+positron flux~\cite{LATCRE, LATCRE1}.
  
We used two independent methods to estimate the residual proton contamination remaining after the selection cuts described above.  The first background estimation method uses fits applied to the flight data alone; it does not require simulating the detector.  The transverse size of showers in the calorimeter provides significant discrimination power for separating electromagnetic showers from hadronic interactions.  In this method we use slightly relaxed selection criteria to increase the statistics of both leptons and hadrons and fit the distribution of transverse shower size to statistically estimate the number of signal (and background) events present in the three regions.

\begin{figure}[]
\includegraphics[width=\linewidth]{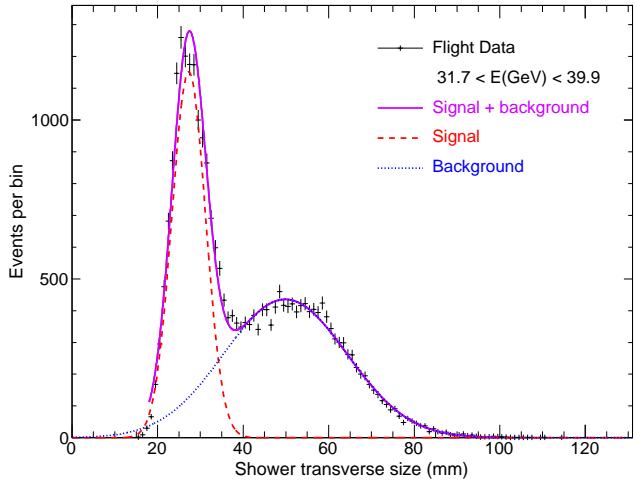}
\caption{Transverse shower size distribution in the electron-only region.  In the positron-only region, the number of events with small transverse shower size is smaller, but the mean and width of the distribution are similar.  See~\cite{supplement} for additional figures.}
\label{fits}
\end{figure}

Distributions of the transverse shower size are produced for each energy bin and for each region.  They are fit with a function that represents the sum of signal and background (Figure~\ref{fits}).  The number of leptons in the sample is determined by integrating the signal function up to 35 mm, where we truncate the integral to reduce the effect of the distribution tail.  The signal and background shower size distributions are each parameterized with a Gaussian.  To evaluate the uncertainty due to small differences between real data distributions and the Gaussian approximation, we applied this fit-based technique to a sample of events collected in survey mode to determine the $e^+ + e^-$ spectrum, and compared it to the previously published LAT measurements~\cite{LATCRE, LATCRE1}.  The spectrum determined with the fit-based background subtraction method agrees with the published one within 5\%. Tests on Monte Carlo simulations confirm that the choice of Gaussian functions is sufficient to obtain the number of signal events at the few-percent level.

The fit is well constrained in the control (both species allowed) region where the number of events is large and the signal-to-background ratio allows us to easily distinguish protons from electrons/positrons.  In the other two regions (electron-only and positron-only) the statistics decrease rapidly with energy and the signal peak parameters are not well constrained.  This is particularly important in the positron region where the background distribution has a larger overlap with the signal.  However, the distribution of shower size depends primarily on particle type (lepton or hadron) and on the angle $\theta$ between the particle track and instrument axis.  The electron and positron regions correspond to particular $\theta$ distributions (the two distributions are very similar to one another) which we reproduce by assembling a  reference set of events from the control region with the same $\theta$ distribution.  We then fix the mean and sigma parameters of the signal Gaussian to the values obtained by fitting this reference data set and fit only the normalization to the data sets from the positron-only and electron-only regions.  No assumptions are made on incoming angle dependence of the background Gaussian, whose parameters are always free.

We evaluated the uncertainty introduced by this fit stabilization procedure by changing the mean and sigma parameters (within $\pm$3 times the uncertainty of the fit values from the reference data sets) and calculating the resulting variation in the number of signal events.  The maximum deviation of the number of signal events is 4\% and is included in the systematic uncertainty for each energy bin.

The second method of estimating the residual proton background is based on a Monte Carlo (MC) simulation.  The LAT detector MC is based on GEANT 4~\cite{Agostinelli2003250} and has been validated using comparisons to flight data and pre-flight beam test data.  We produced a set of 40 billion Monte Carlo protons between 4~GeV and 10~TeV ($\sim$3\% of which pass the trigger and onboard filter) with a power-law distribution in energy with an index of 1.5.  We also simulated 150 million electron events between 4~GeV and 1~TeV ($\sim$13\% of which pass the onboard filter) with a spectral index of 1.  The spacecraft orbit in the simulation matched the actual flight orbit for the data set used.  We re-weight the MC proton flux to the spectrum measured by the AMS-01 experiment (index 2.78)~\cite{Aguilar2002}, and the MC electron flux to the spectrum measured by the Fermi LAT~\cite{LATCRE, LATCRE1} (index 3.08).  We apply the same analysis cuts to the MC and flight data in order to estimate the residual proton contamination.  Roughly 0.1\% of the protons that pass the onboard filter also pass the analysis cuts we designed to reject them.  We traced the trajectories of surviving events and removed events with trajectories blocked by the Earth.  The accuracy of the MC proton rate has been validated by comparing a variety of distributions between MC and flight data.  In particular, we inverted individual cuts in order to produce samples with an enriched proton background contribution.  The rate of MC protons agrees with flight data within $\sim8$\%.


\emph{Results.}  The common systematic uncertainty in the CR electron/positron flux for the two background subtraction methods are: $\pm$5\% effective area; $\pm$5\% onboard filter efficiency in the two lowest energy bins (20.0--31.7~GeV); and $^{+0\%}_{-3\%}$ below 100 GeV, $^{+0\%}_{-10\%}$ above 100 GeV due to atmospheric lepton contamination.  The systematic uncertainty of atmospheric positron contamination is asymmetric because such contamination can only cause us to overestimate, not underestimate, the CR positron flux.  Systematic uncertainties present only for the flight data fitting method are 5--10\% (depending on energy and region) due to discrepancies between the fit shape and the actual distribution and 2--4\% due to using the reference $\theta$ distribution.  Components present only for the MC method are $\sim$8\% due to discrepancies between MC protons and flight data, and 2--10\% CR proton spectral index uncertainty, evaluated using three recent CR proton spectrum measurements (BESS~\citep{Haino200435}, AMS-01~\citep{Aguilar2002}, and PAMELA~\citep{Pamela2011Protons}).  Summing systematic uncertainty components in quadrature (we expect them to be uncorrelated), we estimate 10--16\% (depending on energy) uncertainty in the positron flux for the fit method and 8--19\% for the MC method.  To determine the final uncertainty of each spectral point we add the statistical uncertainty in quadrature; for the MC method this includes both signal and background statistics.

\begin{figure}[]
\includegraphics[width=\linewidth]{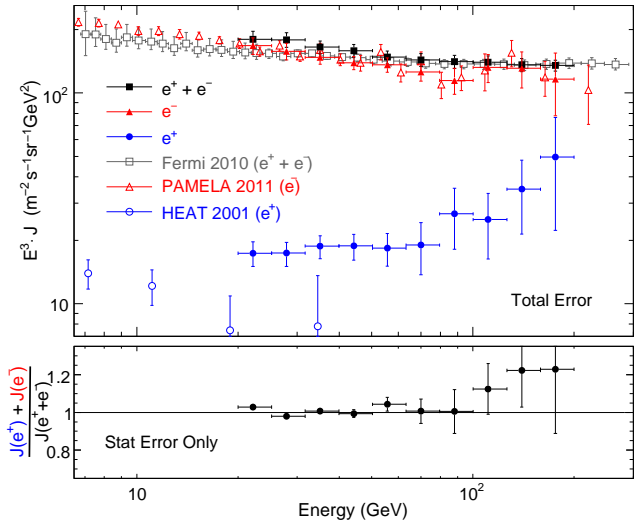}
\caption{Energy spectra for e$^{+}$, e$^{-}$, and e$^{+}$ + $e^{-}$ (control region).  In the control region where both species are allowed, this analysis reproduces the Fermi LAT results reported previously for the total electron plus positron spectrum~\cite{LATCRE, LATCRE1} (gray).  Previous results form HEAT~\cite{HEAT2001} and PAMELA~\cite{Pamela2011CRE} are shown for reference.  The bottom panel shows that the ratio between the sum and the control flux is consistent with 1 as expected.}
\label{fig:ESpecCombine}
\end{figure}

\begin{figure}[]
\includegraphics[width=\linewidth]{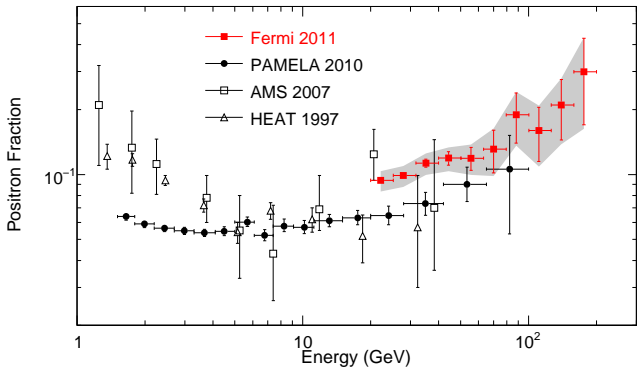}
\caption{Positron fraction measured by the Fermi LAT and by other experiments~\cite{HEAT1997, Aguilar2007, Pamela2}.  The Fermi statistical uncertainty is shown with error bars and the total (statistical plus systematic uncertainty) is shown as a shaded band.}
\label{fig:PosFrc}
\end{figure}

Our best estimates of the spectra are shown in Figure~\ref{fig:ESpecCombine} and Table~\ref{table:fluxtable}.  We chose the fit method for all energy bins except the highest, because this method has slightly smaller uncertainty.  For the highest energy bin we use the MC method because the statistics are not sufficient for fitting.  Results from the MC method are consistent within the total (systematic + statistical) uncertainty of the fit method in all of the bins in the control and electron regions and in seven out of nine bins in the positron region.  The maximum single-bin difference is 1.5 times the total uncertainty.  The positron fraction ($\frac{\mathrm{J(e^{+})}}{\mathrm{J(e^{+})+J(e^{-})}}$) is shown in Figure~\ref{fig:PosFrc}.  Error propagation is performed independently for the fraction and for the individual spectra; the contribution due to effective area uncertainty cancels in the fraction.  The small difference between the combined e$^{+}$ + $e^{-}$ flux we measured in the control region and the total flux we reported previously is due to instrument response functions that have been updated to account for ``ghost events''~\cite{RandoICRC2009}.

\begin{table}[ht]
\begin{tabular}{cccc}
Energy       & J(e$^{+})\times10^{5}$      & J(e$^{-})\times10^{4}$         & $\frac{\mathrm{J(e^{+})}}
									            {\mathrm{J(e^{+})+J(e^{-})}}$  \\
\hline \hline
$20.0-25.2$  & 160$\pm5^{+20}_{-21}$       & 154$\pm1^{+14}_{-14}$          & .094$\pm.003^{+.010}_{-.010}$        \\
$25.2-31.7$  & 80.2$\pm2.9^{+10}_{-10}$   & 72.8$\pm.6^{+6.5}_{-6.5}$      & .099$\pm.003^{+.010}_{-.011}$         \\
$31.7-39.9$  & 43.4$\pm2.0^{+4.9}_{-5.1}$  & 34.1$\pm.4^{+2.5}_{-2.5}$      & .113$\pm.005^{+.012}_{-.012}$        \\
$39.9-50.2$  & 21.8$\pm1.7^{+2.5}_{-2.6}$  & 16.1$\pm.3^{+1.2}_{-1.2}$      & .119$\pm.008^{+.012}_{-.013}$        \\
$50.2-63.2$  & 10.7$\pm1.4^{+1.2}_{-1.3}$  & 7.89$\pm.28^{+.58}_{-.58}$     & .119$\pm.014^{+.012}_{-.013}$        \\
$63.2-79.6$  & 5.52$\pm1.4^{+.66}_{-.68}$  & 3.66$\pm.23^{+.27}_{-.27}$     & .131$\pm.029^{+.014}_{-.014}$        \\
$79.6-100 $  & 3.90$\pm1.2^{+.46}_{-.48}$  & 1.67$\pm.21^{+.12}_{-.12}$     & .189$\pm.049^{+.018}_{-.019}$        \\
$100-126  $  & 1.83$\pm.57^{+.22}_{-.28}$  & .97$\pm.12^{+.08}_{-.08}$      & .160$\pm.045^{+.017}_{-.023}$        \\
$126-159  $  & 1.28$\pm.45^{+.15}_{-.20}$  & .481$\pm.085^{+.039}_{-.039}$  & .210$\pm.065^{+.021}_{-.030}$        \\
$159-200  $  & .911$\pm.48^{+.13}_{-.16}$  & .214$\pm.069^{+.011}_{-.011}$  & .30$\pm.13^{+.03}_{-.04}$            \\
\hline
\end{tabular}
\caption{Flux (GeV$^{-1}$ m$^{-2}$ s$^{-1}$ sr$^{-1}$) and positron fraction as a function of energy (GeV).  Uncertainties are $\pm$stat $\pm$sys.}
\label{table:fluxtable}
\end{table}

The spectrum measured in each of the three regions between 20 and 200 GeV is well described by a power law.  The fit to the positron spectrum is (2.02$\pm$0.22 $\times$ 10$^{-3}$ GeV$^{-1}$ m$^{-2}$ s$^{-1}$ sr$^{-1}$)$(\frac{E}{\mathrm{20 GeV}})^{-2.77\pm0.14}$, while the fit to the electron spectrum is (2.07$\pm$0.13 $\times$ 10$^{-2}$ GeV$^{-1}$ m$^{-2}$ s$^{-1}$ sr$^{-1}$)$(\frac{E}{\mathrm{20 GeV}})^{-3.19\pm0.07}$.  The uncertainties are determined by including the total (statistical plus systematic) uncertainty of each energy bin.  The fitted indices are consistent with the index we reported previously for the total electron plus positron spectrum (3.08$\pm$0.05)~\cite{LATCRE,LATCRE1}.


\emph{Conclusion.}  We measured the CR positron and electron spectra separately between 20 and 200~GeV, using a novel separation technique which exploits the charge-dependent displacement of the Earth's shadow due to the geomagnetic field.  While the positron fraction has been measured previously up to 100 GeV~\cite{Adriani2009,Pamela2} and the absolute flux has been measured previously up to 50~GeV~\cite{HEAT2001, AMS2008}, this is the first time that the absolute CR positron spectrum has been measured above 50 GeV and that the fraction has been determined above 100~GeV.  We  find that the positron fraction increases with energy between 20 and 200~GeV, consistent with results reported by PAMELA~\cite{Adriani2009,Pamela2}.  Future measurements with greater sensitivity and energy reach, such as those by AMS-02~\cite{AMS02}, are necessary to distinguish between the many possible explanations of this increase.


The Fermi LAT Collaboration acknowledges support from a number of agencies and institutes for both development and the operation of the LAT as well as scientific data analysis. These include NASA and DOE in the United States, CEA/Irfu and IN2P3/CNRS in France, ASI and INFN in Italy, MEXT, KEK, and JAXA in Japan, and the K.~A.~Wallenberg Foundation, the Swedish Research Council and the National Space Board in Sweden. Additional support from INAF in Italy and CNES in France for science analysis during the operations phase is also gratefully acknowledged.

\bibliography{PositronPaper}



\addtolength{\textheight}{1cm}

\section{Measurement of separate cosmic-ray electron and positron spectra with the Fermi Large Area Telescope. Supplementary online material.}

\subsection{Shower transverse size distribution}
\begin{figure*}[]
\includegraphics[width=0.44\textwidth]{FitExample_atm60_bin2_electrons.eps}%
\hspace{2em}%
\includegraphics[width=0.44\textwidth]{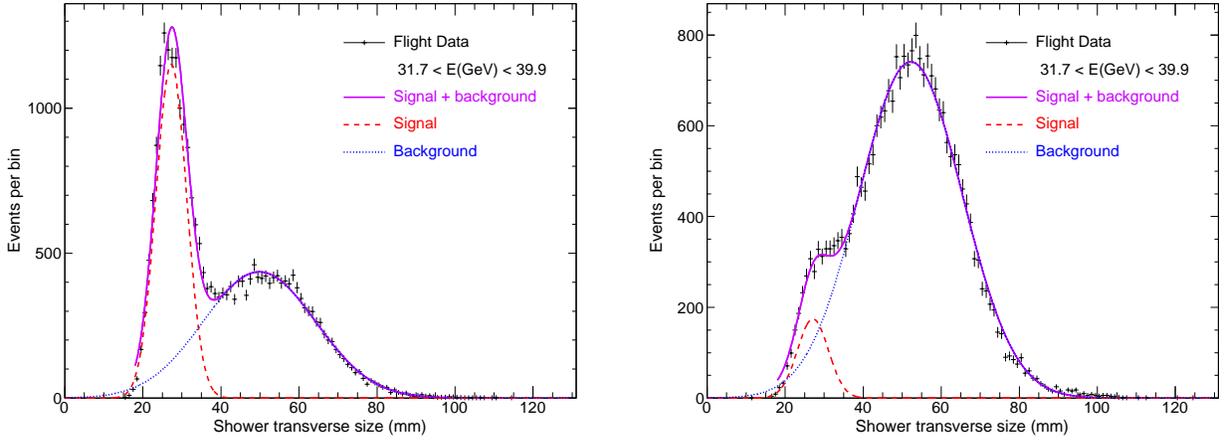}
\caption{Transverse shower size distribution in the electron-only region 
(left panel) and positron-only region (right panel) for the energy interval 
from 31.7 GeV to 39.9 GeV.  \label{fit_bin2}}
\end{figure*}

\begin{figure*}[]
\includegraphics[width=0.44\textwidth]{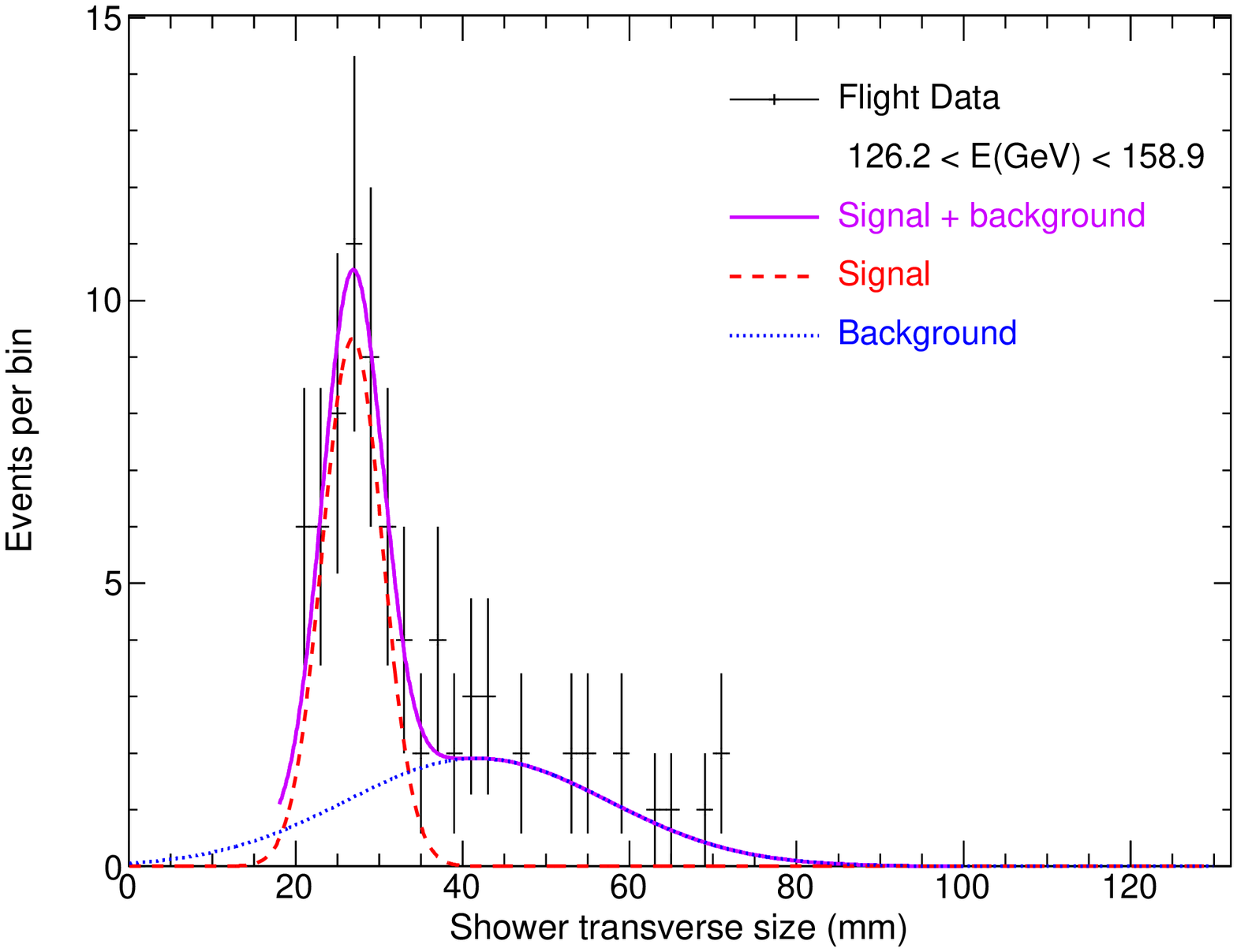}%
\hspace{2em}%
\includegraphics[width=0.44\textwidth]{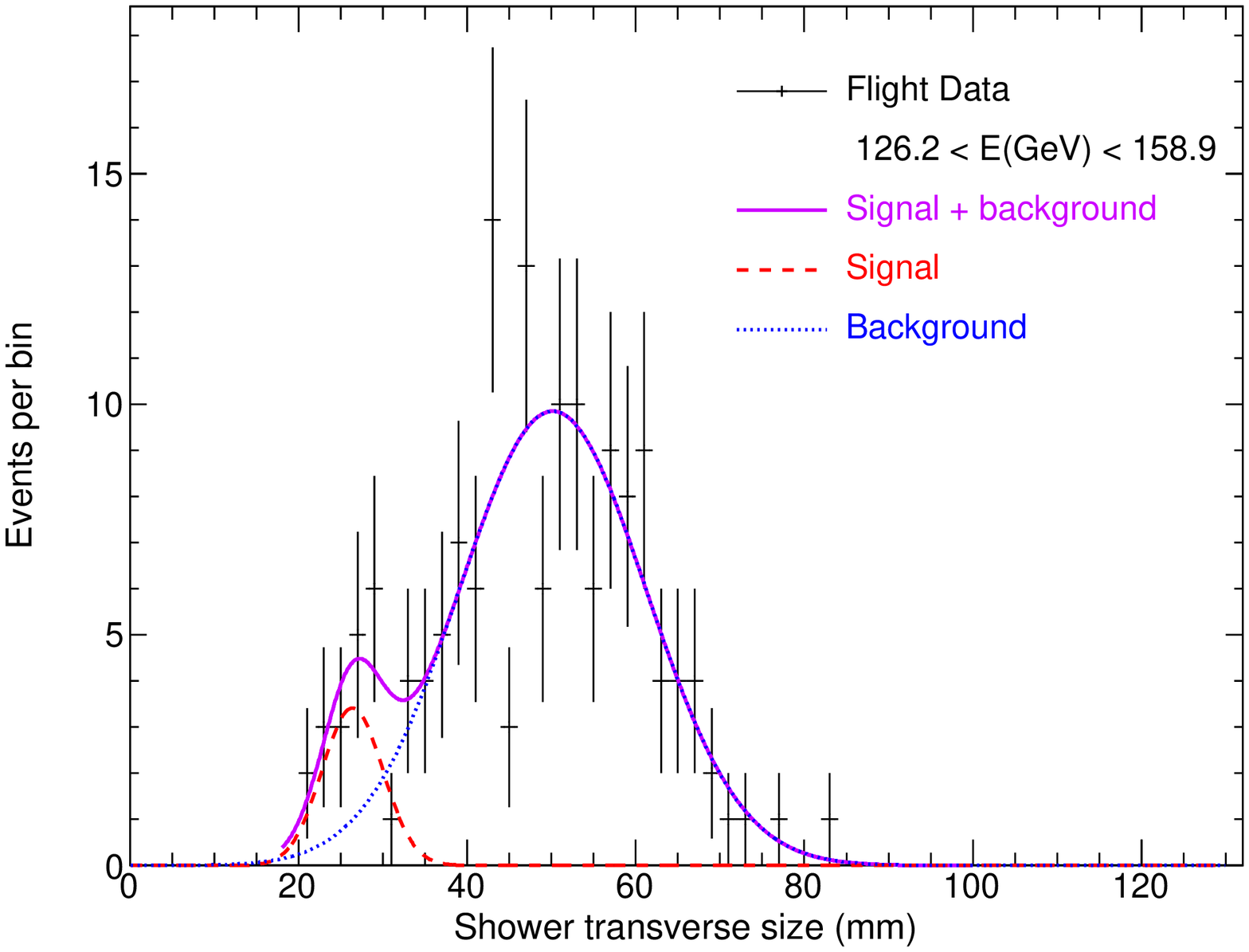}
\caption{Transverse shower size distribution in the electron-only region 
(left panel) and positron-only region (right panel) for the energy interval 
from 126.2 GeV to 158.9 GeV.  \label{fit_bin8}}
\end{figure*}

\begin{figure*}[]
\includegraphics[width=0.44\textwidth]{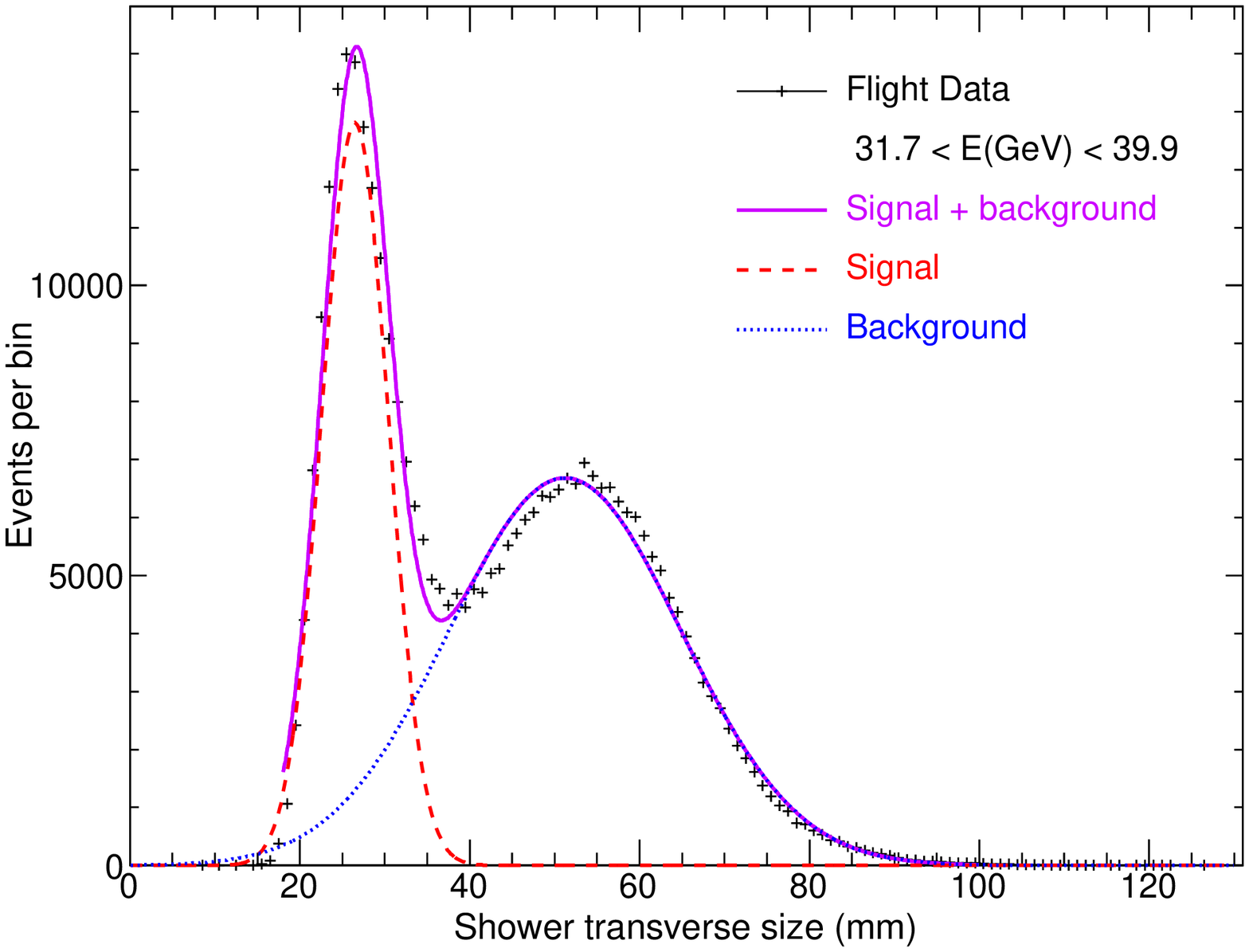}%
\hspace{2em}%
\includegraphics[width=0.44\textwidth]{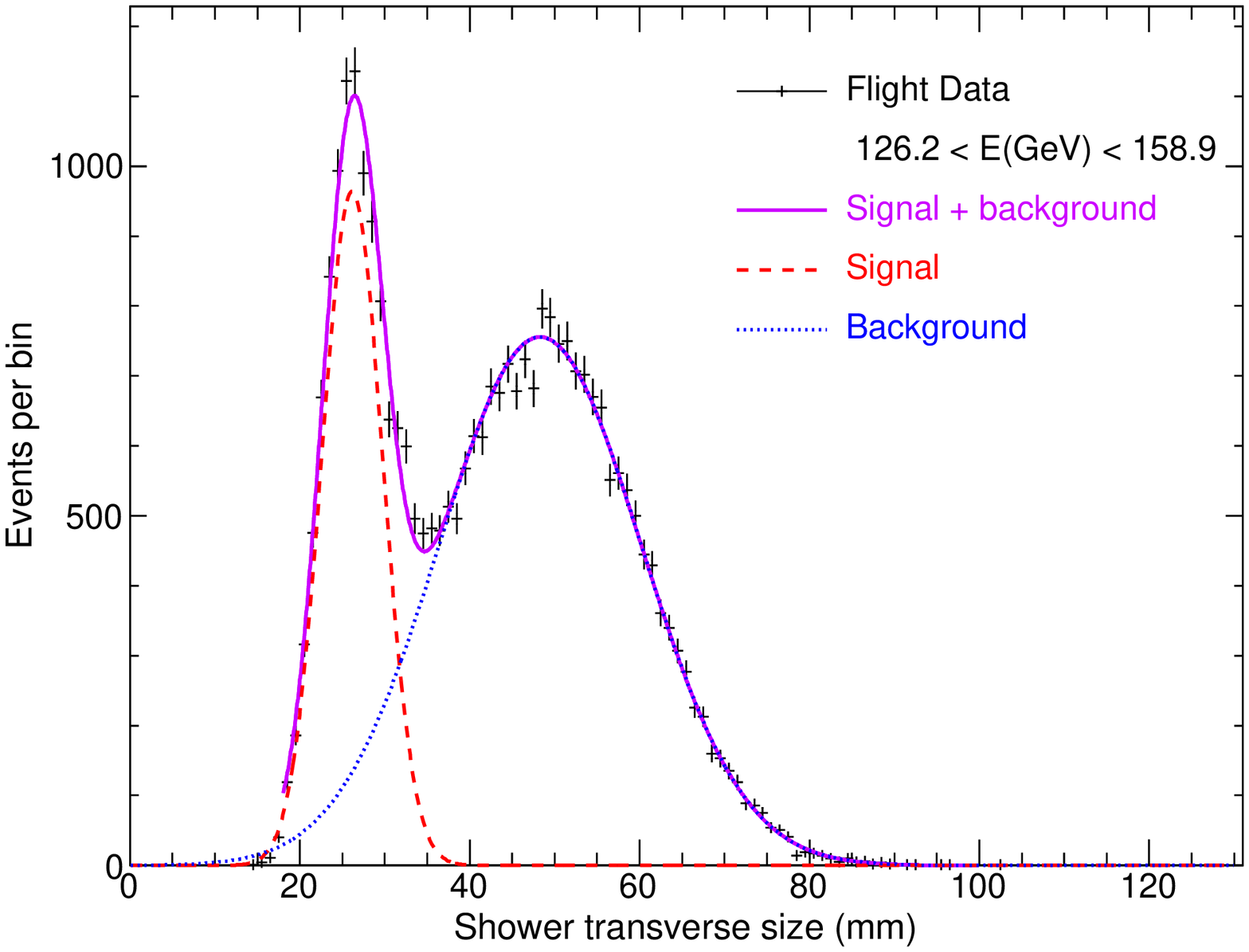}
\caption{Transverse shower size distribution in the control region 
in the two energy intervals \mbox{31.7 -- 39.9~GeV}  (left panel) and \mbox{126.2 -- 158.9~GeV} (right panel).\label{fit_both}}
\end{figure*}

The distribution of the transverse size of showers in the calorimeter 
is used to determine the number of signal events in the three regions used in 
this analysis: the electron-only, the positron-only, and 
the control region (where both species are permitted).
In the paper we used the electron-only case between 31.7 and  39.9 GeV
to explain the technique.
Here we show also the positron-only case in Figure~\ref{fit_bin2}.
Figure~\ref{fit_bin8} shows another example in the energy interval from 126.2 GeV to 158.9 GeV, 
which is the highest energy bin in which the fitting technique is applied.
The distributions of events in the control region for both energy intervals are shown in Figure~\ref{fit_both}.

\end{document}